\documentclass[aps, pre, onecolumn,showpacs,12pt,preprintnumbers,showkeys]{revtex4-1}
\usepackage{amsfonts}

\usepackage{graphicx}
\usepackage{dcolumn}
\usepackage{bm}
\include{biblio}
\begin{document}

\title{Electric double layer of colloidal particles in salt-free concentrated suspensions including 
 non-uniform size effects and orientational ordering of water dipoles}

\author{Jun-Sik Sin}
\email{js.sin@ryongnamsan.edu.kp}
\affiliation{College of Physics, \textbf{Kim Il Sung} University, Taesong District, Pyongyang, DPR Korea}
\author{Kwang-Il Kim}
\affiliation{College of Physics, \textbf{Kim Il Sung} University, Taesong District, Pyongyang, DPR Korea}
\author{Kuk-Chol Ri}
\affiliation{College of Physics, \textbf{Kim Il Sung} University, Taesong District, Pyongyang, DPR Korea}
\author{Dok-Yong Ju}
\affiliation{College of Physics, \textbf{Kim Il Sung} University, Taesong District, Pyongyang, DPR Korea}
\author{Nam-Hyok Kim}
\affiliation{College of Physics, \textbf{Kim Il Sung} University, Taesong District, Pyongyang, DPR Korea}
\author{Chung-Sik Sin}
\affiliation{College of Physics, \textbf{Kim Il Sung} University, Taesong District, Pyongyang, DPR Korea}

\begin{abstract}

The response of a suspension under a variety of static or alternating external fields strongly depends on the equilibrium electric double layer that surrounds the colloidal particles in the suspension. The theoretical models for salt-free suspensions can be improved by incorporating non-uniform size effects and orientational ordering of water dipoles neglected in previous mean-field approaches, which are based on the Poisson-Boltzmann approach. Our model including non-uniform size effects and orientational ordering of water dipoles seems to have quite a promising effect because the model can predict the phenomena like a heavy decrease in relative permittivity of the suspension and counterion stratification near highly charged colloidal particle. In this work we numerically obtain the electric potential, the counterions concentration and the relative permittivity around a charged particle in a concentrated salt-free suspension corrected by non-uniform size effects and orientational ordering of water dipoles. The results show the  worth of such corrections for medium to high particle charges at every particle volume fraction. We conclude that non-uniform size effects and orientational ordering of water dipoles are necessary for the development of new theoretical models to study non-equilibrium properties in concentrated colloidal suspensions.

\end{abstract}


\maketitle

\section{Introduction}
A suspension of charged particles  under static or alternative electric fields is one of the most interesting topics of electrokinetics \cite%
{Lyklema_monograph_1995, Hunter_monograph_1995, Masiliyah_monograph_2006, OBrien_fluidmech_1990, Dukhin_langmuir_1999}. In particular, study of salt-free suspensions has been continuously grown in recent years from experimental and theoretical point of view because in salt-free suspensions, colloidal crystals can be formed at relatively low particle volume fraction than in a suspension with added external salt \cite%
 {OBrien_fluidmech_1990, Dukhin_langmuir_1999, Sood_monograph_1991, Medebach_chemphys_2003}. A salt-free suspension is composed of charged colloidal particles and the added counterions released by the particles in the liquid medium, where the electroneutrality is preserved.

The unique behaviours of salt-free suspensions can be understood by using physical quantities like electrophoretic mobility, electrical conductivity and dielectric response, which are electrohydrodynamic properties of suspension in electrolyte solution  \cite%
{Carrique_physchemb_2006, Carrique_langmuir_2008, Arroyo_langmuir_2008, Ruiz_physchemc_2007, Ruiz_physchemb_2008}. The physical quantities can be obtained by solving electrohydrodynamic equations with initial and boundary conditions.  All of static or dynamic quantities are closely related to the properties of the equilibrium electric double layer surrounding the particles. If the salt-free suspension is dilute, for typical cases, particle-particle electrohydrodynamic interactions can be negligible. However, in the high regime of particle volume fraction, the response to external fields is affected by the interactions which give rise to the mathematical difficulty related with many-body interactions as well as the numerical problems for solving the equations without any approximation.

To overcome this problem, the authors of \cite%
{Carrique_physchemb_2009, Ruiz_physchemb_2009, Ruiz_colloidinterf_2010, Carrique_physchemb_2010} introduced a cell model approach, and the approach successfully was applied to dealing with electrokinetic and rheological properties in concentrated suspension. It was confirmed in \cite%
{Denton_physmatter} that up to moderately strong electrostatic couplings, the cell model accurately predicts osmotic pressures of deionised suspensions in agreement with Monte Carlo simulations and renormalized-effective interaction approaches. In spite of the help of the cell model, because the theoretical models are based on Poisson-Boltzmann(PB) approach, the models maintain the shortcomings of Poisson-Boltzmann approach. 

In \cite%
{Lopez_colloidinterf_2007, Lopez_colloidinterf_2008, Aranda_colloidinterf_2009, Rascon_colloidinterf_2009, Lopez_physchemb_2010}, the authors investigated dilute colloidal suspensions by classical PB equation, which neglects ionic correlation, ionic size effects and solvent polarization. They confirmed the fact that considering the distance of closest approach to the charged particle is important for explaining the overcharging mechanism near the charged particle without explicit consideration of ionic correlation.

Roa and coworkers \cite%
{Roa_pccp_2011_1, Roa_pccp_2011_2, Roa_pccp_2011_3, Roa_colloidinterf_2012} considered the finite sizes of ions by extending Borukhov's approach \cite%
{Borukhov_jpolym_2004} to concentrated salt-free suspension with the help of the cell model. The consideration of finite ion sizes yielded reasonable counterions concentration profiles and electric potential profiles and allowed one to understand  reasonably electrophoretic mobility, electric conductivity and dielectric response. 

To well understand properties of electric double layer, different computational approaches such as Monte Carlo and numerical solutions of integral equations were introduced, but they are unable of making feasible predictions out of equilibrium and involve complex calculations \cite%
{Ibarra_pccp_2009, Gonzalez_physchem_1989}.

In the case of electrolytes, electric double layer theory has a  lengthy history compared to salt-free suspension, dating back to Helmholtz \cite%
{Helmholtz_annphys_1879}. 
The original PB approach proposed by Gouy and Chapmann does not consider the finite volumes of ions, and neglects ionic correlations in electrolyte \cite%
{Gouy_jphys_1910, Chapman_philomag_1913}. In fact, for typical situations including low to medium particle charges and monovalent ions in electrolyte solution, this approach seems to be quite reasonable for representation of the equilibrium problem. The limitations of PB approach yield unphysical ionic concentration profiles near highly charged interfaces and also inability to predict the overcharging phenomena. 

To eliminate such a shortcoming of the PB approach, various modifications of PB approach were pioneered and successfully applied to the practical problems. In order to account for ionic size effect, Stern \cite%
{Stern_zelechem_2007} modified the PB approach considering the finite size effect of ions by combining the Helmholtz model \cite%
{Helmholtz_annphys_1879} with the Gouy-Chapmann mode l\cite%
{Chapman_philomag_1913}. Bikerman \cite%
{Bikerman_philosmag_1942} empirically extended Boltzmann distribution by correcting ion concentration for the volume excluded by ions. In the last two decades, researchers considered finite volumes of ions and water molecules within lattice statistical mechanics approach \cite%
{Borukhov_prl_1997, Borukhov_electrochim_2000, Iglic_physfran_1996,  Bohnic_electrochimica_2001, Bohnic_bioelechem_2002, Bohnic_bioelechem_2005}.  However, most of them are based on the assumption that different species of ions in electrolytes have equal size. Recent studies  \cite%
{Tresset_pre_2008, Chu_biophys_2007, Biesheuvel_colloidinterf_2007,  Li_pre_2011, Li_pre_2012, Boschitsch_jcomchem_2012, Kornyshev_physchemb_2007, Popovic_pre_2013} indicate that difference in sizes of ions enables one to understand phenomena like the asymmetric differential electric capacitance and the stratification of counterions.

It is well known that near a charged surface in an electrolyte the relative permittivity of the electrolyte solution varies according to the distance from the surface \cite%
{Onsager_amchem_1936, Kirkwood_chemphys_1939}. Although the Booth model \cite%
{Booth_chemphys_1951, Booth_chemphys_1955} is widely used in many practices, it has the drawback that the model does not take into account the sizes of both ions and water molecules in electrolyte solution. In \cite%
{Iglic_bioelechem_2010, Gongadze_bioelechem_2012, Gongadze_elechem_2013}, the authors developed water polarization model including size effects of ions and water molecules. The permittivity model of electrolytes well represents the fact that the permittivity of an electrolyte solution may be strongly decreased by orientational ordering of water dipoles and depletion of water molecules.

Recently, we \cite%
{Sin_electrochimica_2015} have incorporated not only the non-uniform size effects of ions and water molecules in electrolyte but also the orientational ordering of water dipoles into the Poisson-Boltzmann approach.
It is evident that in the case of salt-free suspension as well as electrolytes, such effects will influence equilibrium and non-equilibrium properties of electric double layer.

Summarizing, in this study, we will focus on the influence of orientational ordering of water dipoles and non-uniform size effects on the electrostatic properties of electric double layer of a charged particle in a concentrated salt-free suspension.  In other words, we extends the Poisson-Boltzmann approach developed for aqueous electrolytes by us \cite%
{Sin_electrochimica_2015} to the case of salt-free suspension. As in \cite%
{Aranda_colloidinterf_2009, Rascon_colloidinterf_2009, Roa_pccp_2011_1}, our model also incorporates an excluded region in contact with the particle of a hydrated radius size, which yields more realistic representation of solid-liquid interface and also predicts results in good agreement with experimental electrokinetic data. Solving numerically new equations, we will analyze the equilibrium surface potential, relative permittivity of the suspension and counterions concentration profiles inside a cell upon changing particle volume fraction, particle surface charge density and size of the counterions. In order to show the importance of the orientational ordering of water dipoles, the results will be compared with ones of \cite%
{Roa_pccp_2011_1} that takes into account only finite ion size.

\section{Theory}

\subsection{The cell model}

\begin{figure}
\includegraphics[width=1\textwidth]{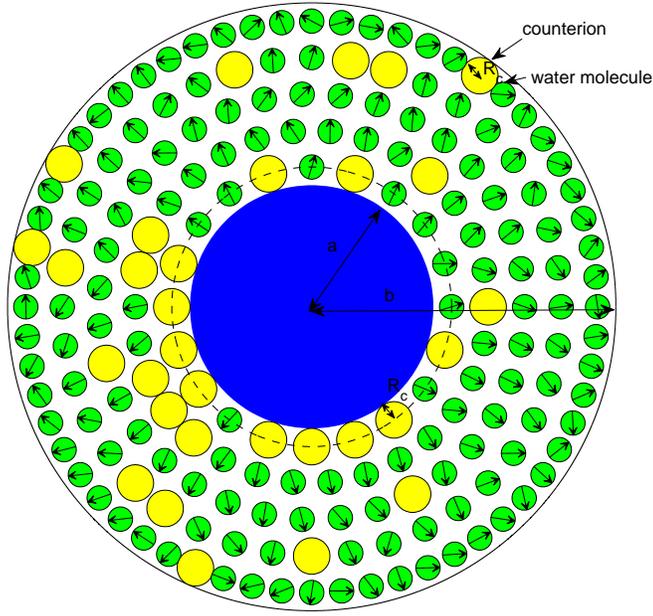}
\caption{(Color online) Cell model including the distance of closest approach of the counterions to the particle surface.}
\label{fig:1}
\end{figure}
Different electrokinetic and rheological phenomena in salt-free concentrated suspensions have been studied by a cell model.  In the cell model, each spherical particle of radius $a$ is surrounded by a concentric shell of the liquid medium, having an outer radius $b$ such that the particle/cell volume ratio in the cell is equal to the particle volume fraction throughout the entire suspension, that is,
\begin{eqnarray}
	\phi=\left(\frac{a}{b}\right)^3.
\label{eq:1}
\end{eqnarray}
 In the cell model, it is assumed that the macroscopic properties of a suspension can be obtained from appropriate averages of local properties in a unique cell. We consider a spherical charged particle of radius  $a$ and surface charge density $\sigma$   immersed in a salt-free medium with the added counterions of valence $z_{c}$. The spherical coordinate system $\left(r,\theta,\phi \right)$     is used and its origin is at the center of the particle. When any external field does not exist, a spherically symmetrical charge distribution surrounds the particle.

\subsection{ Non-uniform size effects and orientational ordering of water dipoles}

The total free energy of the cell can be written in terms of the local electrostatic potential $\psi\left({\bf r}\right)$, the counterions concentration $n_{r}\left({\bf r}\right)$  and the number density of water molecules $n_{w}\left(r\right)=\left<\rho\left(\omega, r \right)\right>_{\omega}$.

 \begin{eqnarray}
	F=\int{d{\bf r}}\left(-\frac{\varepsilon_{0}\varepsilon E^2}{2}+e_{0}z_{c}\psi n_{c}+\left<\rho\left(\omega\right)\gamma{p_{0}}E\cos\omega\right>_{\omega}-\mu_{c}n_{c}-\left<\mu_{w}\left(\omega\right)\rho\left(\omega\right)\right>_{\omega}-Ts\right),
\label{eq:2}
\end{eqnarray}
where $\left<f\left(\omega\right)\right>_{\omega}=\int f\left(\omega\right)2\pi\sin\omega  d\omega$ in which $\omega$ is the angle between the vector {\bf p} and the vector -{\bf E}.  Here {\bf p} is the dipole moment of water molecules and {\bf E} is the electric field strength. The first term is the self energy of the electrostatic field, where $\varepsilon$ equals $n^2$, where $n=1.33$ is the refractive index of water. The next term corresponds to the electrostatic energy of the counterions in the electrostatic mean field, where $e_{0}$ is the elementary charge.
 The third one represents the electrostatic energy of water dipoles\cite%
{Gongadze_bioelechem_2012}, where  $\gamma=\left(2+n^2\right)/2$,  $p_{0}=\left|{\bf p}\right|$ and $E=\left|{\bf E}\right|$. The next two terms couple the system to a bulk reservoir, where $\mu_{c}$ is the chemical potentials of the counterions and $\mu_{w}\left(\omega\right)$ is the chemical potential of water dipoles with orientational angle $\omega$.
$T$ is the temperature and $s$ is the entropy density.

Consider a unit volume of the cell  around the charged particle. The entropy density is the logarithm of the number of translational and orientational arrangements of non-interacting   $n_{c}$ counterions and  $\rho\left(\omega_{i}\right)\Delta\Omega_{i}\left(i=1 \cdots N\right)$ water molecules, where $\Delta\Omega_{i}=2\pi  sin\omega_{i} \Delta\omega$  is an element of a solid angle and $\Delta\omega=\pi/ N$. The counterion and water molecule occupy volumes of $V_{c}$ and $V_{w}$, respectively.
We assume that the volumes are independent of the ionic concentrations.
Within a lattice statistics approach each particle in the suspension occupies more than one cell of a lattice as in \cite%
{Popovic_pre_2013, Sin_electrochimica_2015}. Unlike in \cite%
{Popovic_pre_2013}, orientational ordering of water dipoles as well as translational arrangements of counterions is explicitly considered. We first place $n_{c}$  counterions of the volume $V_{c}$. Accounting for orientational ordering of water dipoles, we put in $\rho\left(\omega_{i}\right) \left(i=0,1,...\right)$ water molecules of the volume $V_{w}$  in the lattice.
 The number of arrangements $W$ is written as 
 \begin{equation}
	W=\frac{n_{s}\left(n_{s}-1\cdot v_{v}\right)\cdots\left(n_{s}-\left(n_{c}-1\right)v_{c}\right)}{n_{c}!}\frac{\left(n_{s}-n_{c}v_{c}\right)\cdots v_{w}}{\lim_{N\rightarrow\infty}\prod^{N}_{i=1}\left(\rho\left(\omega_{i}\right)\Delta\Omega_{i}\right)!},
\label{eq:3}
\end{equation}
where $v_{c,w}= V_{c,w}/a^3$ are the numbers of cells that the counterion and water molecule occupy, respectively.  $n_{s}=1/a^3$ is the number of cells per unit volume and $a$ denotes the linear dimension of one cell.   
Expanding the logarithms of factorials using Stirlings formula, we obtain the expression for the entropy density, $s=k_{B}\ln W$, 
\begin{eqnarray}
	\frac{s}{k_{B}}=\ln W=n_{c}\ln{\frac{V_{c}}{a^3}}+\frac{1}{V_{c}}\ln{\frac{1}{V_{c}}}-n_{c}\ln{n_{c}}-\left(\frac{1}{V_{c}}-n_{c}\right)\ln{\left(\frac{1}{V_{c}}-n_{c}\right)}+\left(\frac{1-n_{c}V_{c}}{V_{w}}\right)\ln{\frac{V_{w}}{a^3}}\nonumber\\+\left(\frac{1-n_{c}V_{c}}{V_{w}}\right)\ln{\left(\frac{1-n_{c}V_{c}}{V_{w}}\right)} -\lim_{N\rightarrow\infty}\sum^{N}_{i=1}{\rho\left(\omega_{i}\right)\Delta\Omega_{i}}\ln{\left(\rho\left(\omega_{i}\right)\Delta\Omega_{i}\right)},
\label{eq:4}
\end{eqnarray}
where $k_{B}$  is the Boltzmann constant.
The variation of the Lagrangian with respect to $n_{c}$ yields an equation from which $n_{c}$   and $\rho\left(\omega\right)$  in the suspension can be obtained.
All lattice cells are occupied by either counterions or water molecules, therefore
\begin{eqnarray}
	n_{s}=n_{c}v_{c}+n_{w}v_{w}.
\label{eq:5}
\end{eqnarray}
Using the method of undetermined multipliers, the Lagrangian of the cell is
\begin{eqnarray}
	L=F-\int\lambda\left({\bf r}\right)\left(1-n_{c}V_{c}-n_{w}V_{w}\right)d{\bf r},
\label{eq:6}
\end{eqnarray}
where $\lambda$ is a local Lagrange parameter.

The Euler$-$Lagrange equations for the Lagrangian are obtained and solved with respect to the functions $n_{c}$ and $\rho\left(\omega\right)$.
The variation of the Lagrangian with respect to $n_{c}$  yields an equation from which $n_{c}$ and $\rho\left(\omega\right)$ in the suspension can be obtained.
 \begin{eqnarray}
	\frac{\delta L}{\delta n_{c}}=e_{0}z_{c}\psi-\mu_{c}-k_{B}T\left(\ln \frac{V_{c}}{a^3}+\frac{V_{c}}{V_{w}}\ln{a^3}+\ln \frac{\left(1-n_{c}V_{c}\right)^{1-V_{c}/V_{w}}}{n_{c}V_{c}}\right)+\lambda V_{c}=0.
\label{eq:7}
\end{eqnarray}
The first boundary condition is $\psi\left(r=b\right)=0$, that fixes the origin of the electric potential at $r = b$. As shown in \cite%
{Roa_pccp_2011_1}, it can be checked that the concentration of counterions, $n_{c}$, is the same irrespective of the origin of the electric potential. Other boundary conditions are $n_{c}\left(r=b\right)=n_{0}$ and $\lambda\left(r=b\right)=\lambda_{0}$,
where $n_{0}$ and $\lambda_{0}$ are unknown coefficients that represent the ionic concentration and the Lagrange parameter where the electric potential is zero, respectively.
Using the boundary conditions we get the chemical potential for counterions from Eq.(\ref{eq:7}):
 \begin{eqnarray}
	\mu_{c}=-k_{B}T\left(\ln \frac{V_{c}}{a^3}+\frac{V_{c}}{V_{w}}\ln{a^3}+\ln \frac{\left(1-n_{0}V_{c}\right)^{1-V_{c}/V_{w}}}{n_{0}V_{c}}\right)+\lambda_{0} V_{c}.
\label{eq:8}
\end{eqnarray}
Inserting Eq.(\ref{eq:8}) into Eq.(\ref{eq:7}), we obtain $n_{c}$   by exponentiation:
 \begin{eqnarray}
	\exp \left(-e_{0}z_{c}\psi/k_{B}T\right)\exp \left(V_{c}\left(\lambda-\lambda_{0}\right)/k_{B}T\right)=\frac{n_{c}\left(1-n_{0}V_{c}\right)^{1-V_{c}/V_{w}}}{n_{0}\left(1-n_{c}V_{c}\right)^{1-V_{c}/V_{w}}}.
\label{eq:9}
\end{eqnarray}
Like the derivation of Eq.(\ref{eq:9}), the expressions for  $\rho\left(\omega\right)$ are simply obtained:
 \begin{eqnarray}
	\rho\left(\omega\right)=\rho_{0}\exp\left(-\frac{\gamma p_{0}E\cos\omega}{k_{B}T}\right)\exp \left(V_{w}\left(\lambda-\lambda_{0}\right)/k_{B}T\right).
\label{eq:10}
\end{eqnarray}
 In general, within our approach the counterions concentration and the number density of water molecules are obtained implicitly not explicitly.

In the case when the counterions and water molecules have the same sizes, that is,  when $V_{w}=V_{c}$, we can recover the result of \cite%
{Gongadze_bioelechem_2012} in Eqs.(\ref{eq:9},\ref{eq:10}). When we neglect orientational ordering of water dipoles, our approach is identical to that of \cite%
{Popovic_pre_2013}.

The Euler$-$Lagrange equation for  $\psi\left(r\right)$  yields the Poisson- Boltzmann equation
\begin{eqnarray}
	\frac{1}{r^2}\frac{d}{dr}\left(r^2\varepsilon_{0}\varepsilon_{r}\frac{d\psi}{dr}\right)=-e_{0}z_{c}n_{c},	
\label{eq:11}
\end{eqnarray}
where
 \begin{eqnarray}
	\varepsilon_{r} \equiv n^2+\frac{\left|{\bf P}\right|}{\varepsilon_{0} E}.
\label{eq:12}
\end{eqnarray}
 Here, ${\bf P}$ is the polarization vector due to a total orientation of point-like water dipoles. From the spherical symmetry of this problem, one can see that the electric field strength is perpendicular to the surface of the charged particle and have the same magnitude at all points equidistant from the surface.  Consequently, ${\bf P}$ is given as \cite%
{Gongadze_bioelechem_2012}
\begin{eqnarray}
	{\bf P}\left(r\right)=c_{w}\left(r\right)\left(\frac{2+n^2}{3}\right)p_{0}\mathcal{L}\left(\gamma{p_{0}}E\beta \right)\hat{{\bf e}}, 
\label{eq:13}
\end{eqnarray}
where a function $\mathcal{L}\left(u\right)=\coth\left(u\right)-1/u$ is the Langevin function and $\hat{{\bf e}}={{\bf E}/E}$ .

Differentiation of Eqs.(\ref{eq:9}, \ref{eq:10}, \ref{eq:5})with respect to the distance from the charged surface provides linear algebraic equations in terms of $d{n_{c}}/dr,d{n_{w}}/dr, dg/dx$, where $g\equiv\left(\lambda-\lambda_{0}\right)/k_{B}T$:
 \begin{eqnarray}
	\frac{d{n_{w}}}{dr}=n_{w}\left[\mathcal{L}\left(\gamma{p_{0}}E\beta\right)\left(\gamma{p_{0}}\beta\right)\frac{dE}{dr}+V_{w}\frac{dg}{dr}\right], 
\label{eq:14}
\end{eqnarray}
 \begin{eqnarray}
	\frac{d{n_{c}}}{dr}=\frac{n_{c}\left(1-n_{c}V_{c}\right)}{1-n_{c}V_{c}^2/V_{w}}\left(\frac{e_{0}z_{c}}{k_{B}T}\frac{d\psi}{dr}+V_{c}\frac{dg}{dr}\right), 
\label{eq:15}
\end{eqnarray}
 \begin{eqnarray}
	\frac{dn_{c}}{dr}V_{c}+\frac{dn_{w}}{dr}V_{w}=0
\label{eq:16}
\end{eqnarray}

Solving the system of Eqs.(\ref{eq:14}),(\ref{eq:15}), (\ref{eq:16}) for $d{n_{c}}/dr,d{n_{w}}/dr,dg/dr$  results in the following coupled differential equations:
 \begin{eqnarray}
	\frac{dn_{c}}{dr}=-\frac{n_{cc}n_{w}V_{c}V_{w}}{n_{w}V_{w}^2+n_{cc}V_{c}^2}\left[[\mathcal{L}\left(\gamma{p_{0}}E\beta\right)\left(\gamma{p_{0}}\beta\right)\frac{dE}{dr}+\frac{V_{w}}{V_{c}}\frac{e_{0}z_{c}}{k_{B}T}\frac{d\psi}{dr}\right]
\label{eq:17}
\end{eqnarray}
 \begin{eqnarray}
	\frac{dn_{w}}{dr}=-\frac{V_{c}}{V_{w}}\frac{dn_{c}}{dr}
\label{eq:18}
\end{eqnarray}
, where $n_{cc}=n_{c}\left[1-\frac{n_{c}V_{c}\left(V_{c}/V_{w}-1\right)}{1-n_{c}V_{c}}\right]^{-1}$.
 
The electrostatic potential, the counterions concentration and the number density of water molecules are obtained by solving Eqs.(\ref{eq:11}), (\ref{eq:17}), (\ref{eq:18}), respectively.

\subsection{ Excluded region in contact with the particle}

	As  Lopez-Garcia {\it et al}.\cite%
{ Lopez_physchemb_2010} and Roa {\it et al}.\cite%
{Roa_pccp_2011_1} did, we account for a distance of closest approach of the counterions to the particle surface, resulting from their finite size. We consider counterions as spheres of radius  $R_{c}$ with a point charge at its center. Counterions can not approach closer to the surface of the particle than their effective hydration radius, $R_{c}$. Consequently, the ionic concentration will be zero in the region between the spherical surface,  $r=a$, and the spherical surface, $r=a+R_{c}$.
The electric potential  $\psi\left(r\right)$  is  determined by the stepwise equation:
 \begin{eqnarray}
	{\frac{d^2\psi\left(r\right)}{dr^2}+\frac{2}{r}\frac{d\psi\left(r\right)}{dr}=0,    a\leq r \leq a+R_{c}}\nonumber\\
	{ Eq. \left(11\right),	 a+R_{c}\leq r \leq b}.
\label{eq:19}
\end{eqnarray}

To completely specify the problem the electric potential should be forced to be continuous at the surface $r=a+R_{C}$, and also its first derivative, which is related to the continuity of the normal component of the electric displacement at that surface. The boundary conditions needed  for solving the problem are
 \begin{eqnarray}
	\psi ' \left(r=a\right)=-\frac{\sigma}{\varepsilon\left(r=a\right)},  \nonumber\\
	\psi\left(r=b\right)=0, \psi ' \left(r=b\right)=0, \nonumber\\
	\psi\left(r=\left(a+R_{c}\right)_{-}\right)=\psi\left(r=\left(a+R_{c}\right)_{+}\right), \psi ' \left(r=\left(a+R_{c}\right)_{-}\right)=\psi ' \left(r=\left(a+R_{c}\right)_{+}\right),
\label{eq:20}
\end{eqnarray}
,where subscript - refers to the region $  a\leq r \leq a+R_{c}$, and subscript +  refers to the region  $a+R_{c}\leq r \leq b$.

\section{Results and Discussion}

\begin{figure}
\includegraphics[width=1\textwidth]{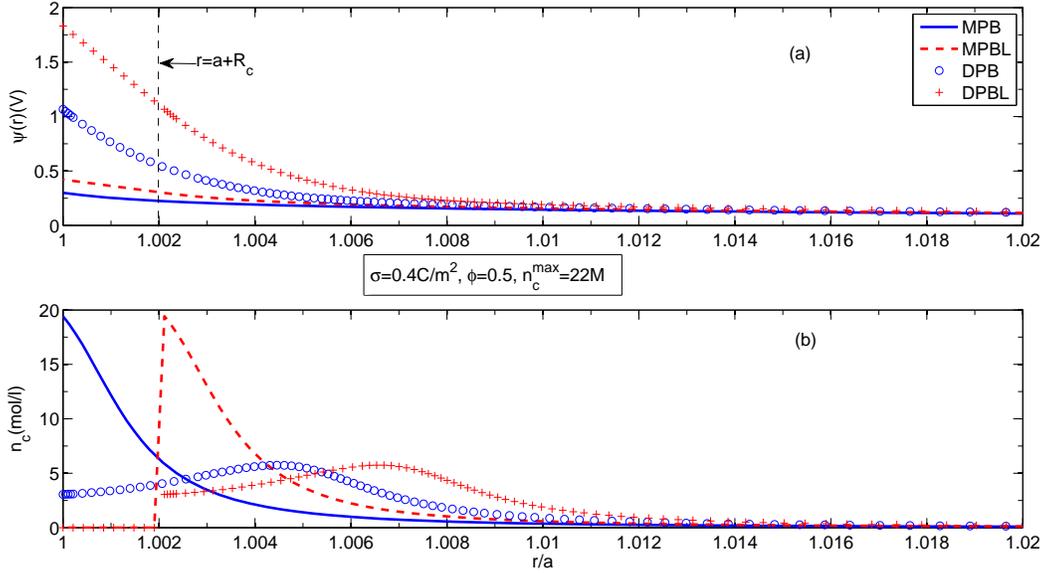}
\caption{(Color online)  Electric potential distribution (a) and counterions concentration (b) along the cell, considering or not orientational ordering of water dipoles and/or the excluded region in contact with the particle. The particle surface charge density, the particle volume fraction and  the maximum possible concentration of counterions due to the excluded volume effect are $\sigma=$0.4C/$m^{2}, \phi=0.5$ and $n_{c}^{max}=22$M, respectively.}
\label{fig:2}
\end{figure}
\begin{figure}
\includegraphics[width=1\textwidth]{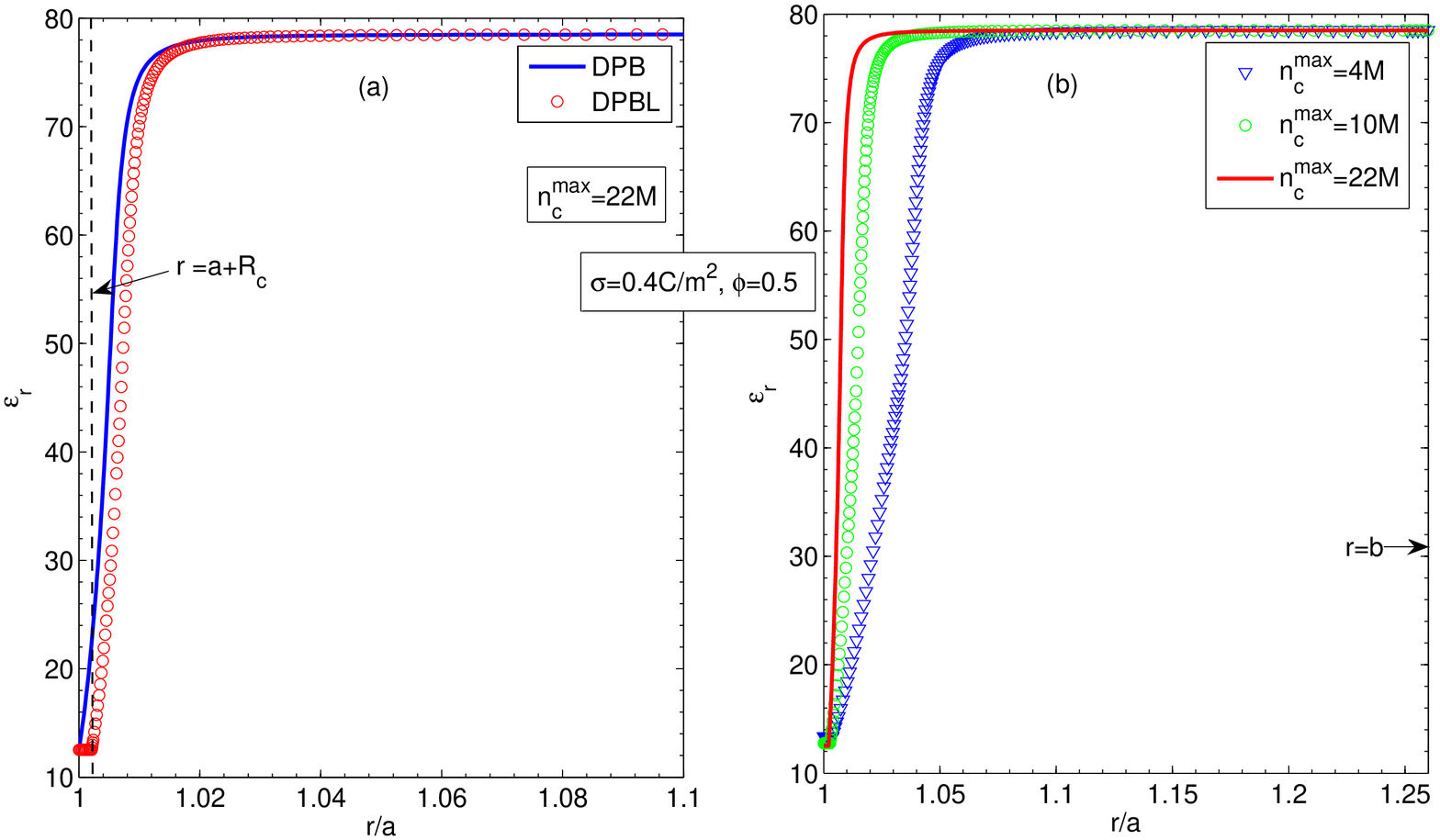}
\caption{(Color online) (a).  Relative permittivity along the cell, considering (circles, DPBL) or not (solid lines, DPB) the excluded region in contact with the particle in the case when orientational ordering of water dipoles is taken into account.   (b).  Relative permittivity along the cell for different ion sizes, considering the excluded region in contact with the particle (DPBL). Triangles, circles and solid lines stand for $n_c^{max}=$ 22M, 10M and 4M. Other parameters are the same as in the FiG. \ref{fig:2}.}
\label{fig:3}
\end{figure}

Under the boundary conditions, we combine Eqs. (\ref{eq:19}, \ref{eq:20}) and solve these differential equations for $n_{c}, n_{w}, \psi$ by using the fourth order Runge-Kutta method, which  is combined with the shooting method.  Calculations are started at the outer surface of the cell. The counterions concentration $n_{0}$ at the surface of the cell is varied to satisfy the boundary condition  Eq.(\ref{eq:20}) for every given surface charge density. For all the calculations, the temperature $T$, the valence of the added counterions $z_{c}$   and  the particle radius $a$ have been taken equal to 298.15K , -1 and 100nm, respectively. As in \cite%
{Gongadze_bioelechem_2012}, the water dipole moment $p_{0}$ should be 3.1D(Debye is $3.336\times10^{-30}$C/m) so that far away from the charged surface($ r=b$) the relative permittivity of the suspension reaches the value of pure water($\epsilon_{r}=78.5$). In calculations, we choose  $n_{0w}=55$M for the number density of water molecules in the bulk suspension.  
In molar concentrations, the values used in the calculations, $n_{c}=$ 22, 10 and 4 M, correspond approximately to counterion effective diameters of $2R_{c}=$0.425, 0.55 and 0.75nm, respectively.

FIG. \ref{fig:2} shows the electric potential distribution, FIG. \ref{fig:2}(a), and the counterions concentration profiles, FIG. \ref{fig:2}(b), along the cell. Solid lines represent the predictions of the equation considering only finite ion sizes without a distance of closest approach to the particle surface (MPB). Dashed lines stand for the results of the equation accounting for finite ion sizes and a distance of closest approach to the particle surface (MPBL). Circles correspond to the outcome of the equation taking into account finite ion sizes and orientational ordering of water dipoles(DPB).  Plus signs exhibit the equation accounting for finite ion sizes, orientational ordering of water dipoles and a distance of closest approach to the particle surface (DPBL). The particle surface charge density have been chosen equal to 0.4C/$m^{2}$   and the particle volume fraction is  $\phi=0.5$(very concentrated suspension), which implies a normalized cell size of $b/a=1.26$. In fact, the authors of \cite%
{Roa_pccp_2011_1} also used the same size of parameters, but did not consider orientational ordering of water dipoles. 

FIG. \ref{fig:2}(a) shows that for our cases, the electric potentials are higher than ones of \cite%
{Roa_pccp_2011_1}. This is attributed to the fact that orientational ordering of water dipoles requires an additional electric energy. In the case when we take into account the distance of closest approach of the counterions to the particle surface( DPBL model), there is an higher increase of the surface potential as in \cite%
{Roa_pccp_2011_1}. Although in the case of DPBL model the surface electric potential reaches 2V that appears unfeasible,  in the case of DPB model the potential is about 1V.  It means that for the case of DPBL, in the region of  $a+R_{c} \leq r\leq b$ the electric potential difference is about 1V, that is below the decomposition potential of water. The rest of potential difference is applied to the region of closest approach to the particle surface. As a result, $\sigma$=0.4C/$m^{2}$ is the maximum possible surface charge density for stable presence of suspension. This is consistent with the fact that in \cite%
{Bohnic_bioelechem_2005, Roa_pccp_2011_1} calculations was performed  within the region of $\sigma \leq 0.4C/m^{2}$, which is called as physiological range. This shows that our approach well represents practical situations. 
We can observe in FIG. \ref{fig:2}(b) that accounting for orientational ordering of water dipoles causes non-monotonous counterions concentration profiles, i.e. the position of the maximum counterions concentration is not located at the position of the surface of the particle. This is due to the statistical mechanical competition effect between orientational ordering of water dipoles and counterion condensation near the particle surface \cite%
{Gongadze_genphysiol_2013, Sin_electrochimica_2015}. For the counterions concentration profiles, our approach mathematically predicts two peaks. The higher one of them is equal to one of the case considering only finite ion size \cite%
{Roa_pccp_2011_1}, $n_{c}^{max}$. The higher peak  occurs at a larger particle surface charge density than corresponding one of the lower peak. The formation of the higher peak requires quite a large particle surface charge density which corresponds to the unphysical value above the decomposition potential of water. 

FIG. \ref{fig:3} shows  the permittivity of the suspension for considering (circles, DPBL) or not (solid lines, DPB) the excluded region in contact with the particle, FIG. \ref{fig:3}(a),  and  for a particle volume fraction $\phi=0.5$ for different ion sizes, FIG. \ref{fig:3}(b), along the cell when orientational ordering of water dipoles is taken into account. The parameters are the same as in FIG. \ref{fig:2}. FIG. \ref{fig:3}(a) shows that the permittivity of the suspension near the particle surface decreases towards the particle surface.  This decrease of $\epsilon_{r}$  is attributed to the increased orientational ordering of water dipoles and  increased depletion of water molecules due to the accumulated counterions near the particle surface. The spatial region with decreased permittivity extends to a larger distance when we take into account the excluded region in contact with the particle.
As shown in FIG. \ref{fig:3}(b), it is obvious that the larger the counterion size, the wider the spatial region with permittivity lower than one of pure water. Eqs. (\ref{eq:12}, \ref{eq:13}) explain that a small number of water dipoles and increased orientational ordering results in decreased permittivity of the suspension, as discussed in FIG. \ref{fig:3}(a). FIG. \ref{fig:3}(a) and FIG. \ref{fig:3}(b) well represent that close to the charge surface the orientation of water molecules and depletion of water molecules result in spatial variation of permittivity \cite%
{Outhwaite_molphys_1976, Butt_physcheminterf_2003,Gongadze_bioelechem_2012}.

\begin{figure}
\includegraphics[width=1\textwidth]{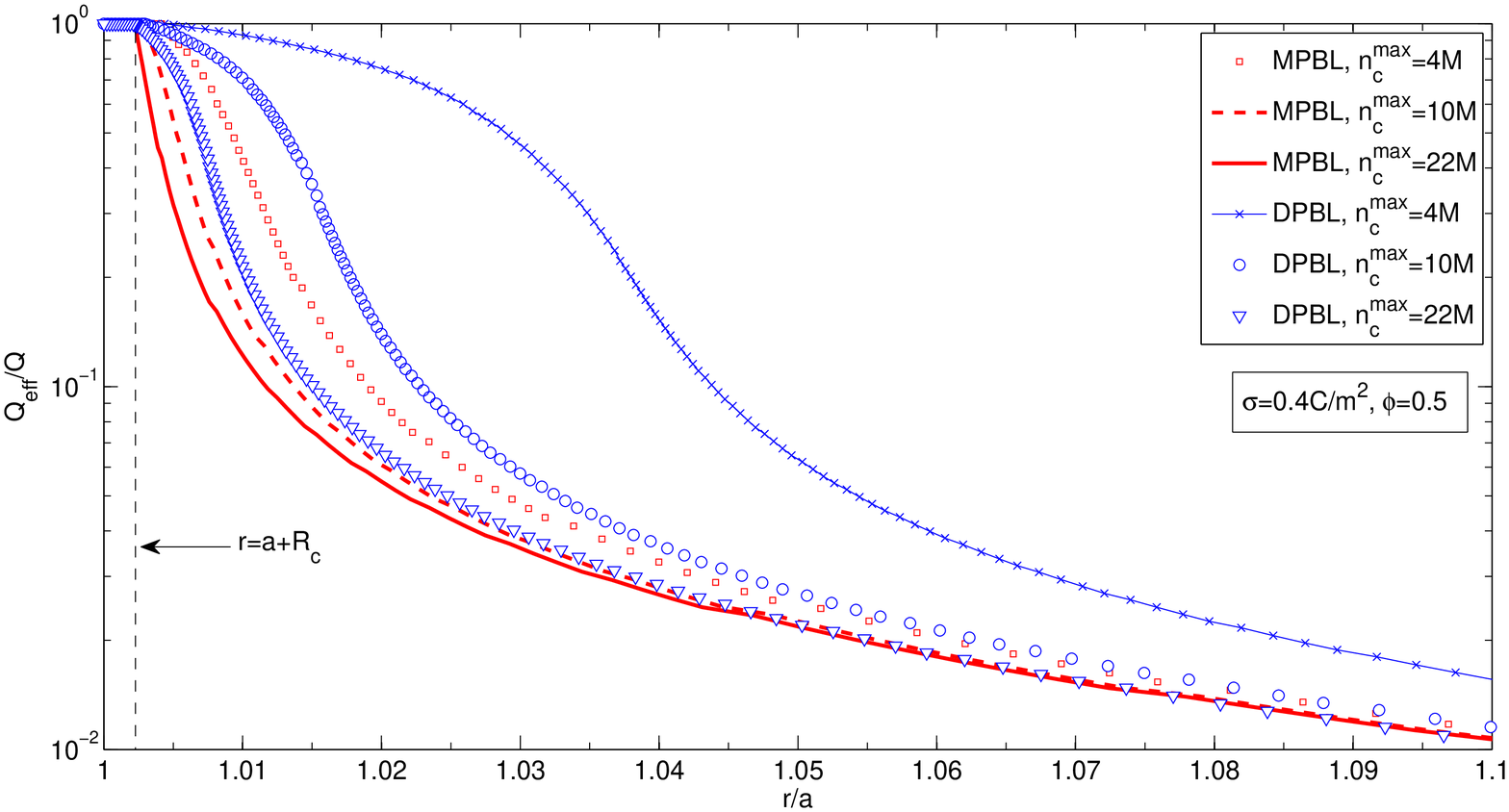}
\caption{(Color online) Effective charge divided by the particle charge along the cell for different ion sizes, considering or not orientational ordering of water dipoles in the case when the excluded region in contact with the particle is taken into account. For the case when orientational ordering of water dipoles is not taken into account(MPBL), squares, dashed lines and solid lines stand for $n_c^{max }=$4M, 10M and 22M, respectively. For the case when orientational ordering of water dipoles is taken into account(MPBL), crosses, circles and triangles correspond to  $n_c^{max }=$4M, 10M and 22M, respectively. Other parameters are the same as in Fig.\ref{fig:3}(b).}
\label{fig:4}
\end{figure}

FIG. \ref{fig:4} shows the effective charge divided by the particle charge along the cell for a particle volume fraction $\phi=0.5$, for different ion sizes and for MPBL and DPBL. The parameters are the same as in Fig. \ref{fig:2}.
It is shown in FIG. \ref{fig:4} that the effective charges for the method(DPBL) studied by considering orientational ordering of water dipoles are larger than corresponding ones of \cite%
{Roa_pccp_2011_1}(MPBL). In other words, orientational ordering of water dipoles diminishes the screening of the suspension.
This is due to the non-monotonous counterions concentration profiles as shown in FIG. \ref{fig:2}(a).
\begin{figure}
\includegraphics[width=1\textwidth]{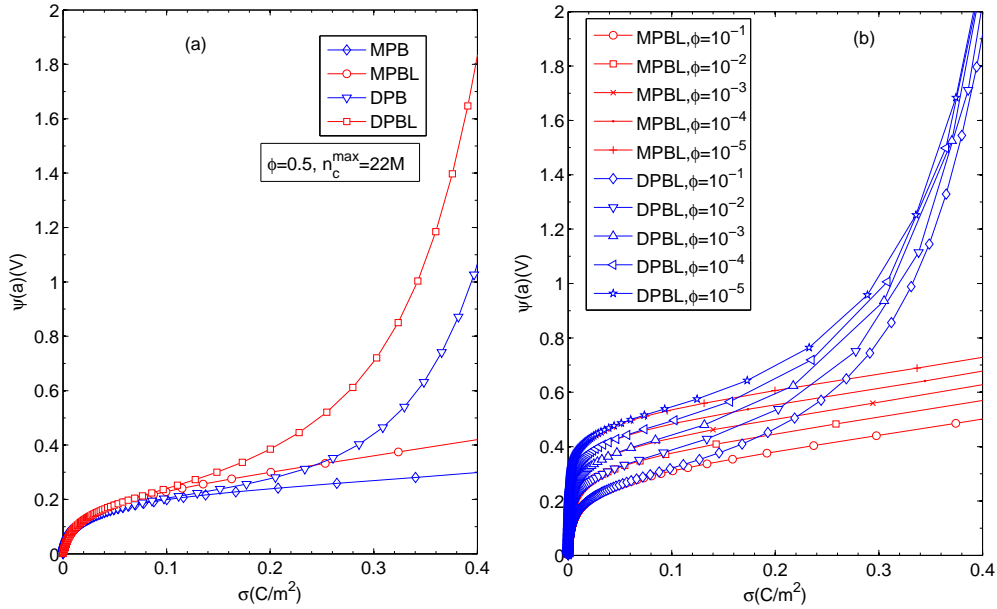}
\caption{(Color online) (a). Surface electric potential for different values of particle surface charge density for different ion sizes, considering or not orientational ordering of water dipoles and/or the excluded region in contact with the particle(MPB, MPBL, DPB, DPBL).
(b). Surface potential against the particle surface charge density for different particle volume fraction values. Other parameters are the same as in Fig. \ref{fig:2}.}
\label{fig:5}
\end{figure}
\begin{figure}
\includegraphics[width=1\textwidth]{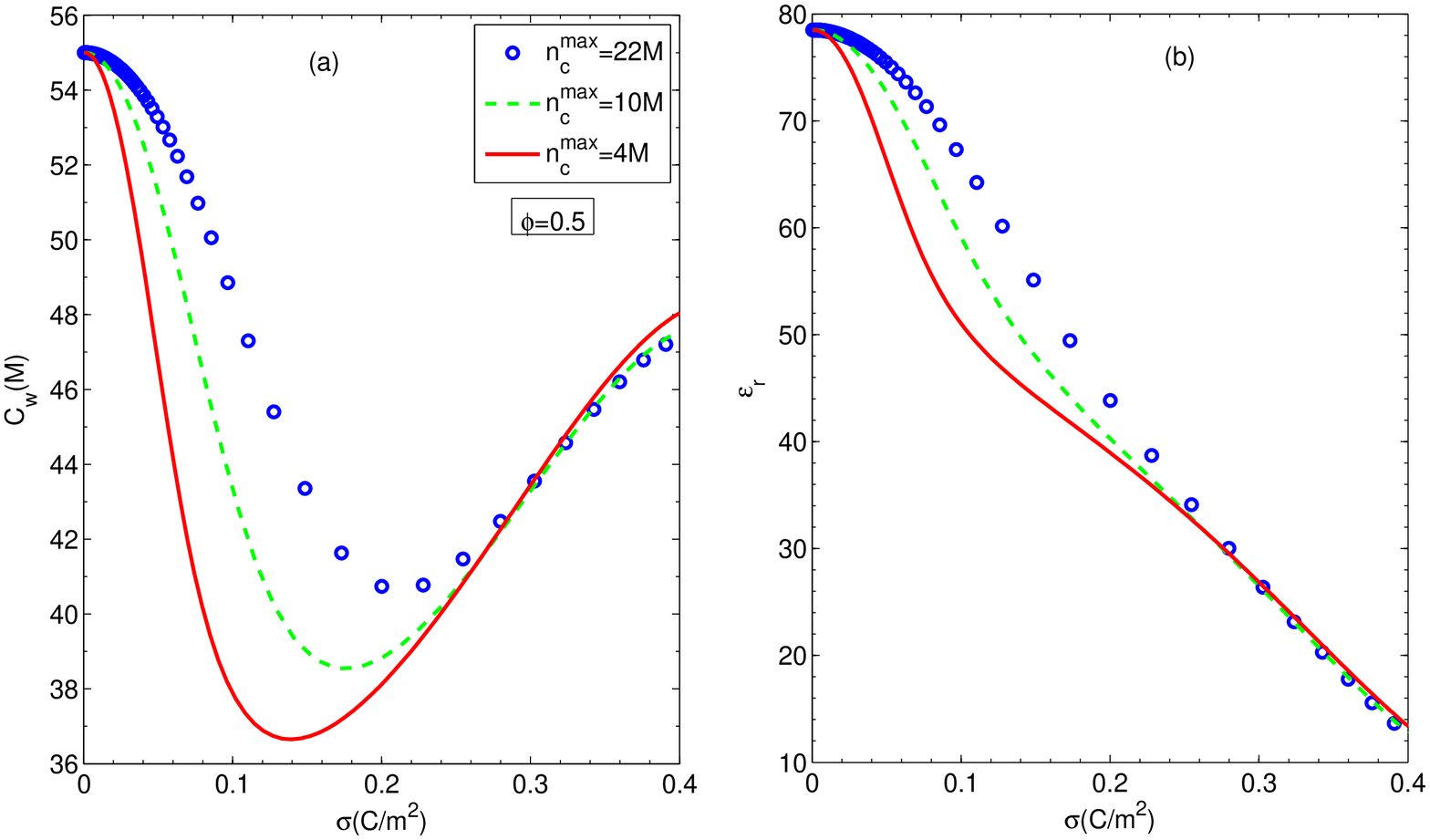}
\caption{(Color online) Surface number density of water molecules (a) and surface relative permittivity (b) as functions of particle surface charge density for different ion sizes, considering the excluded region in contact with the particle and orientaional ordering of water dipoles. Circles, dashed lines and solid lines stand for $n_c^{max}=$22M, 10M and 4M. Other parameters are the same as in Fig. \ref{fig:2}.}
\label{fig:6}
\end{figure}

FIG. \ref{fig:5} displays the equilibrium electric surface potential as a function of particle surface charge density for different ion sizes and  $\phi=0.5$ , FIG. \ref{fig:5}(a), and for different particle volume fractions and $n_{c}^{max}=22$M, FIG. \ref{fig:5}(b). It is shown in FIG. \ref{fig:5}(a) that although in all cases the surface electric potential increases with particle surface charge density, the shapes of the profiles clearly differ from each other. In the cases when orientational ordering of water dipoles is taken into account(DPB and DPBL), the surface potential with particle surface charge density rapidly increases compared to the case of no accounting for water dipoles(MPB and MPBL), respectively.  As shown in  \cite%
{Roa_pccp_2011_1}, PB equation predicts that in the region of low surface charge density, the surface potential sharply increases with the surface charge density. This is in agreement with the predictions of our theory as well as MPB. However, in the region of medium and high charge density, the surface electric potential predicted by MPB increases more rapidly than ones of PB as proved in  \cite%
{Roa_pccp_2011_1}. In the cases of DPB and DPBL, the behavior of the classical condensation effect is severely limited, because in the high regime of the surface charge the additional counterions are located in farther positions form the particle surface than ones predicted by MPB and MPBL as shown in FIG. \ref{fig:2}(b). Consequently, in the case when orientational ordering of water dipoles is explicitly considered, the further increase of the surface potential is attributed to the larger extent of counterion condensation compared to the case of no accounting for orientational ordering of water dipoles. It can be intuitively explained that from the medium regime, the results of DPBL and DPB theories are significantly deviated from corresponding ones of MPBL and MPB, respectively. In fact, a high surface charge density yields a wide spatial region with low permittivity compared to the case of the low surface charge density. Combining the above fact with the boundary condition  $E=\frac{\sigma}{\epsilon_{r}\left(r=a\right)}$ and the definition of electric field strength ${\bf E}= -\frac{d\psi}{dr}$, we can easily verify that the results of DPBL and DPB theories are higher than ones of MPBL and MPB, respectively.
FIG. \ref{fig:5}(b) shows that in the case of orientational ordering of water dipoles is taken into account(DPBL), the surface electric potential increases with the particle volume fraction as in the case of  MPBL. When the particle concentration rises, the space for the counterions inside the cell gets small and, consequently the screening of the particle charge is enhanced, thus reducing the value of the surface potential. This fact will have significant consequences on the electrokinetic properties of colloidal particles in concentrated salt-free suspensions, as shown in \cite%
{Aranda_colloidinterf_2009, Rascon_colloidinterf_2009} for dilute suspensions in electrolyte solutions.

FIG. \ref{fig:6}(a) and FIG. \ref{fig:6}(b) show the number density of water molecules and the relative electric permittivity of the suspension at the particle surface according to the particle surface charge density for different ion sizes, respectively. FIG. \ref{fig:6}(a) exhibits that accounting for simultaneously orientational ordering of water dipoles and finite size effects yields a non-monotonous behavior of the number density of water molecules at the particle surface. It should be noted that the smaller counterion size involves the shallower valley of the number density of water molecules at the particle surface. Then we note that all the number densities of water molecules are nearly equal in the high regime of surface charge density. According to Eqs.(\ref{eq:12}, \ref{eq:13}), the number density of water molecules is proportional to the relative permittivity at any position. On the other hand, the electric permittivity decreases with magnitude of the electric field strength at any position.  Even though in the case of larger number density of water molecules, electric field strength gets slightly smaller, the sequence of magnitude of the electric permittivities for all cases studied is in agreement with one of magnitude of corresponding number densities of water molecules as shown in Fig. \ref{fig:6}(b).

\section{Conclusions}

In this work we have studied the influence of non-uniform size effects and orientational ordering of water dipoles on the description of the equilibrium electric double layer of a spherical colloidal particle in a concentrated salt-free suspension.

We have used a cell model approach to deal with particle-particle interaction, and extended Poisson-Boltzmann equations to account for such non-uniform size effects and orientational ordering of water dipoles.  The theoretical procedure has followed that used by the authors \cite%
{Sin_electrochimica_2015}
 but with the additional inclusion of a region of closest approach for counterions to the particle surface. Unlike in \cite%
{Roa_pccp_2011_1}, our PB approach can consider different sizes of different species of counterions as well as the difference in sizes of counterions and water molecules. We also have calculated the counterions concentration, equilibrium electric potential and relative permittivity for concentrated suspensions. The results have shown that orientational ordering of water dipoles is quite important for medium to high particle charges at every particle volume fraction, and even more if the distance of closest approach of the counterions to the particle surface is taken into account.

This equilibrium model presented in this study will be used as the base of non-equilibrium models for the response of a suspension to external electric fields. Experimental results concerning the static or dynamic electrophoretic mobilities, electrical conductivity and dielectric response, should be compared with the predictions of non-equilibrium models. Although the stratification of counterions associated with the difference in size of the counterions has not been accessed in the present study, consideration of the difference in size of the counterions is very important for realistic salt-free suspensions which contain different species of counterions. Recently, the authors of \cite%
{Li_pre_2011, Li_pre_2012} studied the competition of multiple counterions of different valences and different sizes in binding to the surface of a spherical colloidal particle by both a mean-field theory and Monte Carlo simulations. The both methods predicted the stratification of counterions around the charged surface and found that the ionic valence-to-volume ratios, instead of ionic valences alone, are the key factors that determine the binding of counterions to the charged surface. Because  the authors of \cite%
{Li_pre_2011, Li_pre_2012}  did not consider decreased permittivity of suspension near the charge particle, it will be interesting to study the phenomena by using our method. Additionally,  Monte Carlo approach accounting for both non-uniform size effects and orientational ordering of water dipoles is necessary for studying deeply the phenomena. As a final conclusion, we emphasize that we can apply this model to prediction of non-equilibrium properties in realistic salt-free concentrated suspensions. The task will be performed by the authors in the near future.

\
\end{document}